\documentclass[twocolumn,amsmath,pra]{revtex4}%
\usepackage{amsfonts,amssymb}
\usepackage{graphicx}%

\begin{document}
\title{Quantum properties of a parametric four-wave mixing in a Raman-type atomic system}

\author{A.V. Sharypov$^{1,2}$}
\email[Corresponding author: ]{asharypov@yandex.ru}
\author{Bing He$^3$}
\author{V.G. Arkhipkin$^{1}$}
\author{S.A. Myslivets$^{1}$}
\affiliation{$^1$Kirensky Institute of Physics, Federal Research Center KSC SB RAS, 50, Akademgorodok, Krasnoyarsk 660036, Russia}
\affiliation{$^2$Science Center "Newton Park", 1 Mira, Krasnoyarsk, 660049, Russia}
\affiliation{$^3$Department of Physics, University of Arkansas, Fayetteville, AR 72701, USA}

\begin{abstract}
We present a study of the quantum properties of two light fields used to
parametric four-wave mixing in a Raman type atomic system. The system realizes
an effective Hamiltonian of beamsplitter type coupling between the light
fields, which allows to control squeezing and amplitude distribution of the
light fields, as well as realizing their entanglement. The scheme can be
feasibly applied to engineer the quantum properties of two single-mode light
fields in properly chosen input states.

\end{abstract}
\maketitle


\section{\bigskip Introduction}

Squeezing \cite{SqRevLvovsky} and entanglement \cite{EntRevHorodecki} are two
important features of quantum light and have no counterparts in the classical
framework. They are the cornerstone of quantum computation and quantum
communication with continuous variable light fields, where quantum information
is usually encoded into Gaussian states of light \cite{EntCVBraunstein,
BookQuantumInf}. Entanglement and squeezing are normally generated through
parametric processes such as parametric down conversion
\cite{IntroQOGerryKnight} and four-wave mixing (FWM) \cite{Harris1, Harris2,
Xiao1, Xiao2, Lett}.

FWM can realize an effective coupling of beamsplitter type between two light
fields \cite{SharypovPRA13}, which can be applied to generate the important
categories of photonic quantum states, such as cat states and symmetric
entangled states. In the present work we consider a Raman-type dispersive FWM
to realize a similar beamsplitter type coupling for other applications
including transferring squeezing between different modes, amplifying the
amplitude of one squeezed mode and entangling different quantum modes. There
are two degenerate counterpropagating electromagnetic fields as the pump in
our concerned system, and they realize an effective standing wave creating the
spatial modulation of the nonlinearities for the input quantum light fields.
This structure behaves like photonic crystals that are widely used for control
light propagation \cite{PC1, PC2}. To avoid the influence of the quantum
noises that are critical to the entanglement generation (see, e.g.
\cite{n1,n2,noise}), we use a dispersive parametric interaction, so that the
one-photon detuning and two-photon detuning in the process are large enough to
prevent the real excitation of the atomic system from its ground states and
only the quantum state of the light fields will be changed during the
close-loop parametric interaction, which involves the two counter-propagating
quantum modes and the standing wave of the pump field.

The rest of the paper is organized as follows. We first present a description
of the model in Sec. II. The effective Hamiltonian for the system in the
dispersive regime is derived in Sec. III. In Sec. IV we study the properties
of quantum light fields, such as the amplitude dynamics, squeezing
transferring and entanglement generation, after finding the evolving light
field modes. We summarize the results in Sec. V.

\bigskip

\section{\bigskip The model}

\begin{figure}
[ptb]
\begin{center}
\includegraphics[
height=3.2652in,
width=2.9082in
]%
{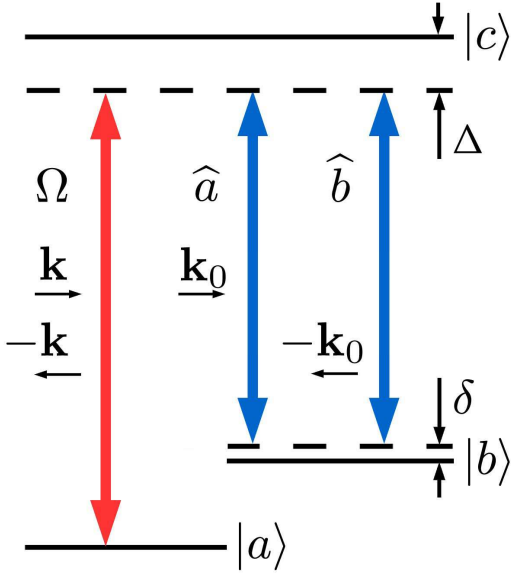}%
\caption{Scheme of a Raman-type dispersive four-wave mixing of two
counterpropogating quantum modes $\widehat{a}$ and $\widehat{b}$ with the wave
vectors $\pm k_{0}$ and the classical standing wave with the Rabi frequency
$\Omega$ in a media of a three-level quantum systems.}%
\label{Fig1Sys}%
\end{center}
\end{figure}

Let us consider a three-level atomic system with the ground state $\left\vert
a\right\rangle $ and two upper states $\left\vert b\right\rangle $ and
$\left\vert c\right\rangle $; see Fig. \ref{Fig1Sys}. Through the system the
input quantum light fields of single mode, the blue ones in Fig.
\ref{Fig1Sys}, effectively interact with one another. The classical standing
electromagnetic field with the Rabi frequency $\Omega$ implements the
transition $\left\vert a\right\rangle -\left\vert c\right\rangle $. Such
classical standing wave is applied along $z$ - direction and can be decomposed
into two running wave with the wave vectors $\pm k$. The two
counterpropagating quantum modes that are coupled to the transition
$\left\vert b\right\rangle -\left\vert c\right\rangle $ with the coupling
parameter $g$\ also propagate along $z$\ direction.%

In the interaction picture the Hamiltonian of the system in Fig. \ref{Fig1Sys}
takes the following form ($\hbar\equiv1$)

\bigskip%
\begin{align}
H  &  =-\Delta\left\vert c\right\rangle \left\langle c\right\vert
-\delta\left\vert b\right\rangle \left\langle b\right\vert \nonumber\\
&  W\left(  \left\vert a\right\rangle \left\langle c\right\vert +\left\vert
c\right\rangle \left\langle a\right\vert \right)  +\widehat{s}^{\dagger
}\left\vert b\right\rangle \left\langle c\right\vert +\widehat{s}\left\vert
c\right\rangle \left\langle b\right\vert , \label{H1}%
\end{align}
\bigskip where
\begin{align*}
W  &  =2\Omega\cos kz\\
\widehat{s}  &  =g\left(  \widehat{a}e^{ik_{0}z}+\widehat{b}e^{-ik_{0}%
z}\right)  .
\end{align*}
In Eq. (\ref{H1}) we introduce the one-photon detuning $\Delta=\omega
-\omega_{ac}$ and the two-photon detuning $\delta=\omega-\omega_{0}%
-\omega_{ab}$, where $\omega$, $\omega_{0}$ are the optical frequencies of the
classical field and quantum modes correspondingly, while $\omega_{ac}$ and
$\omega_{ab}$ are the frequencies of the corresponding transitions.

\section{\bigskip Effective Hamiltonian}

In what follows we will consider a regime of the dispersive interaction, where
the large detunings of the electromagnetic fields prevent the real transitions
of the three-level system from its ground state. So, the atomic system remains
the same initial state $\left\vert a\right\rangle $ during the interaction,
and only the quantum states of the optical fields can be changed. The
classical field is assumed to be much stronger than the quantum ones. In order
to derive the effective Hamiltonian in the dispersive regime we use an
adiabatic elimination technique \cite{SharypovPRA13, s1, GardinerThesis}. We
start from the Schr\"{o}dinger equation $i\hbar\frac{d\Psi\left(  t\right)
}{dt}=H\Psi\left(  t\right)  $ and project it onto atomic states $\left\vert
a\right\rangle $, $\left\vert b\right\rangle $ and $\left\vert c\right\rangle
$:%

\begin{align}
i\frac{d}{dt}\left\langle a|\Psi\left(  t\right)  \right\rangle  &
=W\left\langle c|\Psi\left(  t\right)  \right\rangle \label{S1g}\\
i\frac{d}{dt}\left\langle b|\Psi\left(  t\right)  \right\rangle  &
=-\delta\left\langle b|\Psi\left(  t\right)  \right\rangle +\widehat
{s}^{\dagger}\left\langle c|\Psi\left(  t\right)  \right\rangle \label{S1c}\\
i\frac{d}{dt}\left\langle c|\Psi\left(  t\right)  \right\rangle  &
=-\Delta\left\langle c|\Psi\left(  t\right)  \right\rangle +W\left\langle
a|\Psi\left(  t\right)  \right\rangle +\widehat{s}\left\langle b|\Psi\left(
t\right)  \right\rangle , \label{S1a}%
\end{align}
\bigskip following the derivations detailed in \cite{SharypovPRA13,
GardinerThesis}\ under the conditions ($n$ is the maximal photon number in the
quantum modes) \bigskip%
\begin{align*}
\left\vert \frac{W}{\Delta}\right\vert  &  <<1\text{ and}\\
\frac{g\sqrt{n}W}{\left\vert \Delta\delta\right\vert }  &  <<1,
\end{align*}
which allows to avoid one-photon and two-photon transition so that the system
remains in its ground state $\ \left\vert a\right\rangle $ and only the state
of the light fields will be changed. Then we can eliminate the states
$\left\vert c\right\rangle $ and $\left\vert b\right\rangle $ from the
dynamical equations, to obtain the effective dynamical evolution
\begin{equation}
i\frac{d}{dt}\left\langle a|\Psi\left(  t\right)  \right\rangle =\widetilde
{H}_{eff}\left\langle a|\Psi\left(  t\right)  \right\rangle , \label{S2g}%
\end{equation}

where%

\[
\widetilde{H}_{eff}=\frac{\widehat{s}^{\dagger}\widehat{s}W^{2}}{\Delta
^{2}\delta}%
\]
is the effective Hamiltonian\bigskip.

Furthermore we can simplify the present Hamiltonian and eliminate $z-$
dependence. By omitting the fast oscillation terms and making tranformation
$T_{\delta}=\exp i\Delta k\left(  a^{\dagger}a+b^{\dagger}b\right)  z$, we
obtain a beamsplitter type Hamiltonian%

\begin{equation}
H_{eff}=\chi_{0}\left(  \widehat{a}^{\dagger}\widehat{a}+\widehat{b}^{\dagger
}\widehat{b}\right)  +\sigma_{0}\left(  \widehat{a}\widehat{b}^{\dagger
}+\widehat{a}^{\dagger}\widehat{b}\right)  \label{Heff}%
\end{equation}

where%

\begin{equation}
\chi_{0}=\frac{2\Omega^{2}g^{2}}{\Delta^{2}\delta}-\Delta kc, \label{c1}%
\end{equation}

\bigskip with $\Delta k=k-k_{0}$, is due to a self-phase modulation and%

\begin{equation}
\sigma_{0}=\frac{\Omega^{2}g^{2}}{\Delta^{2}\delta} \label{c22}%
\end{equation}

is the cross-coupling between the quantum modes.

\section{\bigskip Engineering of the quantum light fields}

The field operators' evolution is described by the Heisenberg equation as the
two coupled propagation ones%

\begin{align*}
\frac{d}{dz}\widehat{a}  &  =i\chi\widehat{a}+i\sigma\widehat{b}\\
\frac{d}{dz}\widehat{b}  &  =-i\sigma\widehat{a}-i\chi\widehat{b},
\end{align*}

\bigskip where the parameters $z=\alpha_{0}z_{0}$ (the replacement
$t\rightarrow z_{0}/c$ has been used), $\chi=-\chi_{0}c/\alpha_{0}$ and
$\sigma=-\sigma_{0}c/\alpha_{0}$ are renormalized with the unpertubed
absorption coefficient $\alpha_{0}$, and the counterpropagation geometry of
the quantum modes is taken into account. Since the two modes have a
counterpropagating geometry, the boundary conditions become $\widehat
{a}\left(  z=0\right)  =\widehat{a}_{0}$ and $\widehat{b}\left(  z=L\right)
=\widehat{b}_{L}$, where $L$\ is the length of the medium. The solution for
the output modes $\widehat{a}\left(  z=L\right)  =\widehat{a}_{L}$ and
$\widehat{b}\left(  z=0\right)  =\widehat{b}_{0}$ can be written in the
following form:%

\begin{align}
\widehat{a}_{L}  &  =S_{1}\left(  L\right)  \widehat{a}_{0}+S_{2}\left(
L\right)  \widehat{b}_{L}\label{a1}\\
\widehat{b}_{0}  &  =S_{2}\left(  L\right)  \widehat{a}_{0}+S_{1}\left(
L\right)  \widehat{b}_{L} \label{b1}%
\end{align}

where

$\bigskip$%
\begin{align}
S_{1}  &  =\left[  \cos sL-i\frac{\chi}{s}\sin sL\right]  ^{-1}\label{S1}\\
S_{2}  &  =iS_{1}\frac{\sigma}{s}\sin sL\label{S2}\\
s  &  =\sqrt{\chi^{2}-\sigma^{2}} \label{s}%
\end{align}

$\bigskip$

This result is in agreement with that obtained by treating the currently
concerned system as a classical one \cite{ArkhipkinOptLett}.

During the process the commutation relations, $\left\langle \widehat{a}%
_{L}\widehat{a}_{L}^{\dagger}\right\rangle -\left\langle \widehat{a}%
_{L}^{\dagger}\widehat{a}_{L}\right\rangle =\left\langle \widehat{b}%
_{0}\widehat{b}_{0}^{\dagger}\right\rangle -\left\langle \widehat{b}%
_{0}^{\dagger}\widehat{b}_{0}\right\rangle =\left\vert S_{1}\right\vert
^{2}+\left\vert S_{2}\right\vert ^{2}=1$, for the photon modes are always
preserved as it should be.

An interesting property of the system is that the parameter $s$ can be
imaginary when $\chi^{2}<\sigma^{2}$. This situation can happen in the range
of the parameters%

\begin{equation}
\frac{1}{3}<P<1, \label{BG}%
\end{equation}

where we have introduced a new notation for the dimensionless parameter%

\begin{equation}
P\equiv\frac{\Omega^{2}g^{2}}{\Delta^{2}\delta\left(  \Delta k\right)  c}.
\label{P}%
\end{equation}

This situation is analogous to the presence of the band gap in photonic
crystal \cite{PC1, PC2}, as predicted in \cite{Su2005}.

In the following, we will apply the above-mentioned dynamics to study the
properties of propagation such as the conversion, and squeezing transferring
between the light fields, and entanglement generation under the different
input parameters.

\subsection{field mode swapping}

We start with the discussion of how the amplitude of the quantum modes can be
changed in our concerned process. We can find the photon number in the modes
$A_{a}=\left\langle \widehat{a}^{\dagger}\widehat{a}\right\rangle $ and
$A_{b}=\left\langle \widehat{b}^{\dagger}\widehat{b}\right\rangle $ from
Eqs.(\ref{a1}) and (\ref{b1}):%

\begin{align*}
A_{a}^{L}  &  =\left\vert S_{1}\right\vert ^{2}A_{a}^{0}+\left\vert
S_{2}\right\vert ^{2}A_{b}^{L}\\
A_{b}^{0}  &  =\left\vert S_{1}\right\vert ^{2}A_{b}^{L}+\left\vert
S_{2}\right\vert ^{2}A_{a}^{0},
\end{align*}

where it is assumed that initially the two modes $\widehat{a}$ and
$\widehat{b}$\ are not correlated $\left\langle \widehat{a}_{0}^{\dagger
}\widehat{b}_{L}\right\rangle =\left\langle \widehat{a}_{0}\widehat{b}%
_{L}^{\dagger}\right\rangle =0$. In the numerical calculation we consider the
case when input state for the mode $\widehat{a}$ is in the coherent state
$\left\vert \alpha\right\rangle $ and the second mode $\widehat{b}$ is in the
vacuum state $\left\vert 0\right\rangle $. In this case the amplitudes of
photon modes take a simple form%

\begin{align*}
A_{a}^{L}  &  =\left\vert S_{1}\right\vert ^{2}\left\vert \alpha\right\vert
^{2}\\
A_{b}^{0}  &  =\left\vert S_{2}\right\vert ^{2}\left\vert \alpha\right\vert
^{2}.
\end{align*}

\begin{figure}
[ptb]
\begin{center}
\includegraphics[
height=4.5064in,
width=2.929in
]%
{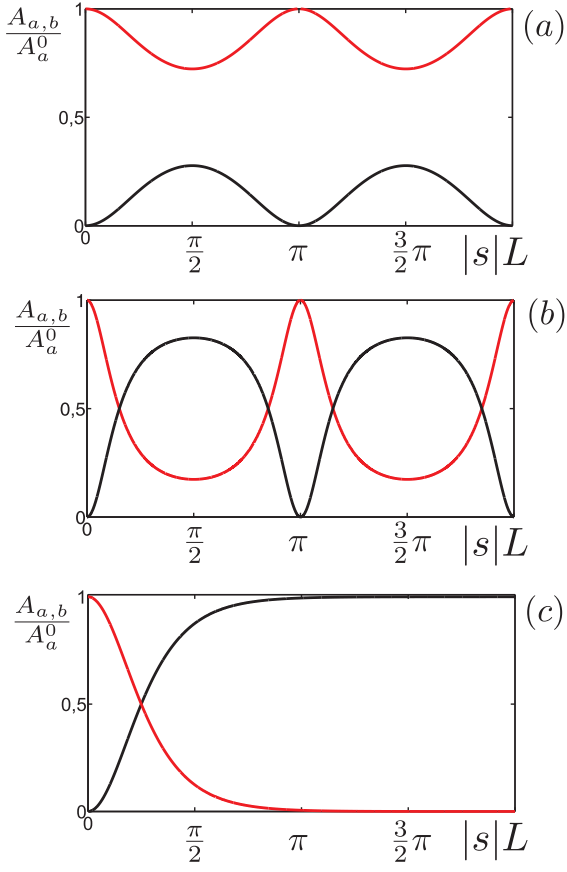}%
\caption{Dynamics of the normalized over $A_{0}=\left\vert \alpha\right\vert
^{2}$\ quantum field's amplitude as a function of the dimensionless parameter
$\left\vert s\right\vert L$\ where $s$\ is given in Eq.\ref{s} and $L$\ is the
interaction length. The red line is the function $A_{a}/A_{a}^{0}$ and the
black one is the $A_{b}/A_{a}^{0}$. a) $P=10$ b) $P=1.1$ c) $P=0.4$}%
\label{fig2amp}%
\end{center}
\end{figure}

The possibility of the squeezing transferring by coherent process involving
electromagnetically induced transparency in atomic system was studied in
\cite{Zubairy1995}, but principally it is impossible to obtain more than 25\%
of the initial squeezing in that way. Here we demonstrate more than 90\% of
the squeezing transferring between the modes. Amplitude of the squeezed mode
and amount of squeezing is easy to control by the intensity of the pump field.

\bigskip

\subsection{\bigskip squeezing transferring}

We define the quadratures of the quantum modes%

\begin{align}
X_{a,b}  &  =\left[  \widehat{a}_{L}\left(  \widehat{b}_{0}\right)
+\widehat{a}_{L}^{\dagger}\left(  \widehat{b}_{0}^{\dagger}\right)  \right]
/2\label{QuadX}\\
Y_{a,b}  &  =\left[  \widehat{a}_{L}\left(  \widehat{b}_{0}\right)
-\widehat{a}_{L}^{\dagger}\left(  \widehat{b}_{0}^{\dagger}\right)  \right]
/2i \label{QuadY}%
\end{align}

and their fluctuations $\left\langle \Delta X_{a,b}\right\rangle
^{2}=\left\langle X_{a,b}^{2}\right\rangle -\left\langle X_{a,b}\right\rangle
^{2}$ and $\left\langle \Delta Y_{a,b}\right\rangle ^{2}=\left\langle
Y_{a,b}^{2}\right\rangle -\left\langle Y_{a,b}\right\rangle ^{2}$.

If the fluctuation in one of the quadratures is less than $1/4$, the field
will be in squeezed state. For the input coherent states in the considered
system, the output are always in coherent states and there is no way to get a
squeezed state%

\[
\left\langle \Delta X_{a,b}\right\rangle ^{2}=\left\langle \Delta
Y_{a,b}\right\rangle ^{2}=\frac{1}{4}\left(  \left\vert S_{1}\right\vert
^{2}+\left\vert S_{2}\right\vert ^{2}\right)  =\frac{1}{4}%
\]

The situation will be different in the case when one of the modes is initially
in a squeezed state. As an example we assume that mode $\widehat{a}$ is in the
coherent state $\left\vert \alpha\right\rangle $ and mode $\widehat{b}$ in the
squeezed state $\left\vert \xi\right\rangle $ with squeezing parameter $r$.
The expressions for the four quadratures fluctuations are%

\begin{align*}
\left\langle \Delta X_{a,b}\right\rangle ^{2}  &  =\frac{1}{4}+\\
&  \frac{1}{4}\left[  2\left\vert S_{2,1}\right\vert ^{2}\sinh^{2}r-\left(
S_{2,1}^{2}+\left(  S_{2,1}^{\ast}\right)  ^{2}\right)  \sinh r\cosh r\right]
\\
\left\langle \Delta Y_{a,b}\right\rangle ^{2}  &  =\frac{1}{4}+\\
&  \frac{1}{4}\left[  2\left\vert S_{2,1}\right\vert ^{2}\sinh^{2}r+\left(
S_{2,1}^{2}+\left(  S_{2,1}^{\ast}\right)  ^{2}\right)  \sinh r\cosh r\right]
\end{align*}

\begin{figure}
[ptb]
\begin{center}
\includegraphics[
height=4.5114in,
width=3.0079in
]%
{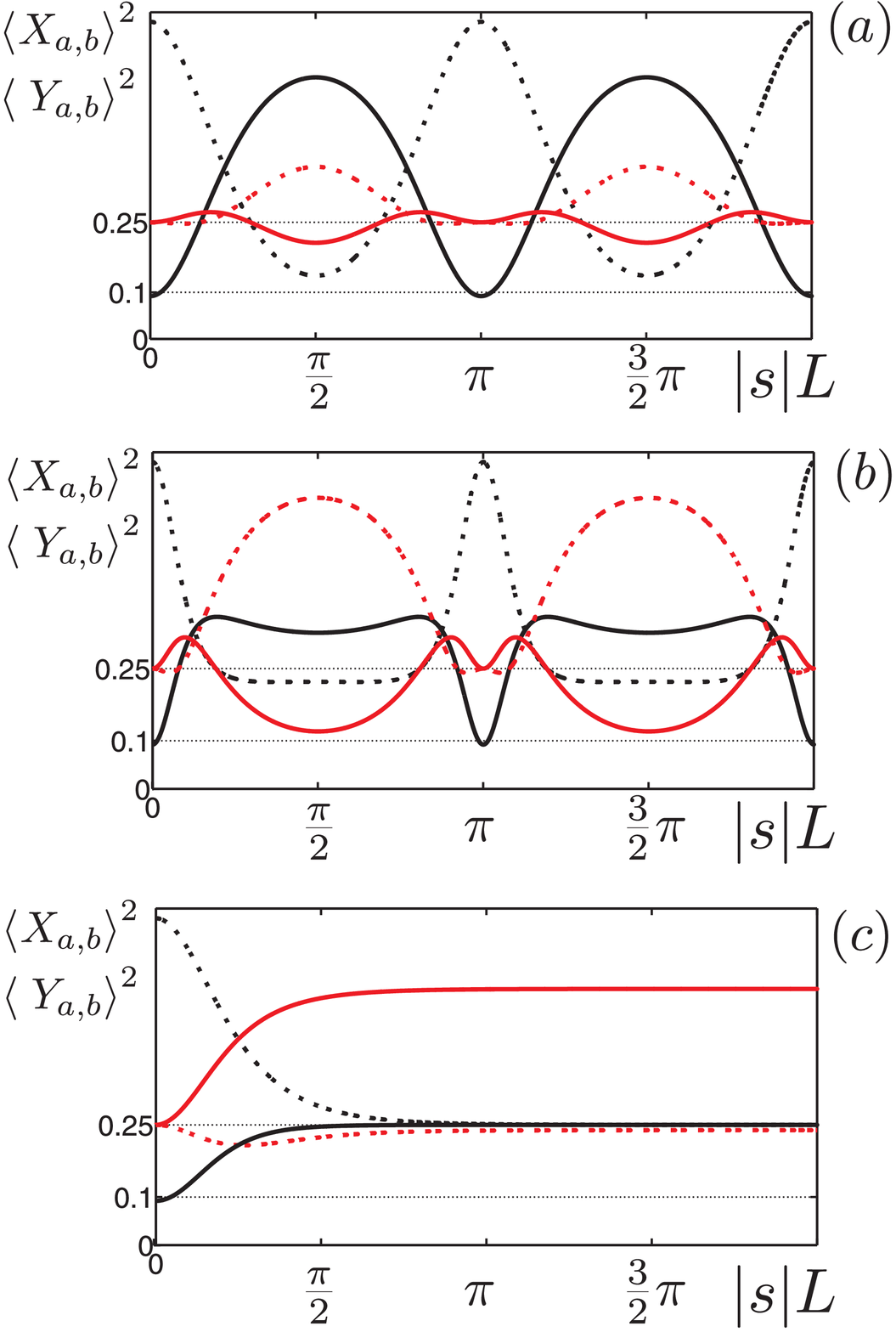}%
\caption{Dynamics of the fields quadratures fluctuations as a function of the
dimentionless parameter $\left\vert s\right\vert L$ where $s$\ is given in
Eq.\ref{s} and $L$\ is the interaction length. $\left\langle \Delta
X_{a}\right\rangle ^{2}$ - red solid line, $\left\langle \Delta X_{b}%
\right\rangle ^{2}$ - black solid line, $\left\langle \Delta Y_{a}%
\right\rangle ^{2}$ - red doted line, $\left\langle \Delta Y_{b}\right\rangle
^{2}$ - black doted line. a) $P=10$ b) $P=1.1$ c) $P=0.4$}%
\label{Fig3Sq}%
\end{center}
\end{figure}

In this case the amplitude of the squeezed mode can be sufficiently amplified
due to the parametric energy transferring from the coherent mode. In Fig.
\ref{Fig3Sq}(a) about $P=10$ [see Eq. (\ref{P})] there is the effective
squeezing transferring between two quadratures of the same mode. At the same
time in Fig. \ref{fig2amp}(a) we can see that at the point $L=\frac{\pi}{2}$
the mode, which is initially weak and squeezed, gets sufficiently amplified
due to the parametric interaction with another quantum mode and preserves
almost the initial level of the squeezing. Next we consider the case when the
squeezing transfers from one mode to another [$P=1.1$ as in Fig. \ref{Fig3Sq}
(b)]. The quantum mode which is initially coherent state and has a large
amplitude quantum mode around the point $L=\frac{\pi}{2}$ will become squeezed
[see Fig. \ref{Fig3Sq}(b)] and it's amplitude will be still sufficiently large
[see Fig. \ref{fig2amp}(b]). For the case when the parameters lie in the
region where the band gap exists $P=0.4$ [see Eq. (\ref{P})], there will be no
sufficient squeezing in any quadrature as in Fig. \ref{Fig3Sq}(c).\bigskip

\bigskip%

The possibility of the squeezing transferring by coherent process involving
electromagnetically induced transparency in atomic system was studied in
\cite{Zubairy1995}, but principally it is impossible to obtain more than 25\%
of the initial squeezing in that way. Here we demonstrate more than 90\% of
the squeezing transferring between the modes. Amplitude of the squeezed mode
and amount of squeezing is easy to control by the intensity of the pump field.

\bigskip

\subsection{\bigskip entanglement generation}

In this section we demonstrate the possibility of generating the entanglement
between quantum modes. As the criteria of entanglement we use the inequality
$Q\equiv\left(  \Delta u\right)  ^{2}+\left(  \Delta\upsilon\right)  ^{2}<1$
\ \cite{EntCreteria}, where $u=X_{a}+X_{b}$ and $\upsilon=Y_{a}-Y_{b}$. The
quadratures $X_{a,b}$ and $Y_{a,b}$ are given in Eqs. (\ref{QuadX}) and
(\ref{QuadY}).\ When both modes are in the coherent state, there will be no
possibility to generate entanglement and function $Q$ is simply larger than
$1$. When mode $\widehat{a}$ is in a coherent state $\left\vert \alpha
\right\rangle $ and mode $\widehat{b}$ in a squeezed state $\left\vert
\xi\right\rangle $, the system will be able to generate an entanglement
between the modes in a certain range of system parameters. For the concerned
system we have%

\begin{align*}
Q  &  =\left(  \Delta u\right)  ^{2}+\left(  \Delta\upsilon\right)  ^{2}\\
&  =\left(  1+\sinh^{2}r\right)  \left(  \left\vert S_{1}\right\vert
^{2}+\left\vert S_{2}\right\vert ^{2}\right) \\
&  -\frac{1}{2}\sinh r\cosh r\left[  S_{1}^{2}+S_{2}^{2}+\left(  S_{1}^{\ast
}\right)  ^{2}+\left(  S_{2}^{\ast}\right)  ^{2}\right]
\end{align*}

Through numerical analysis we find a few regimes where two quantum modes can
be entangled.%

\begin{figure}
[ptb]
\begin{center}
\includegraphics[
height=4.523in,
width=2.7015in
]%
{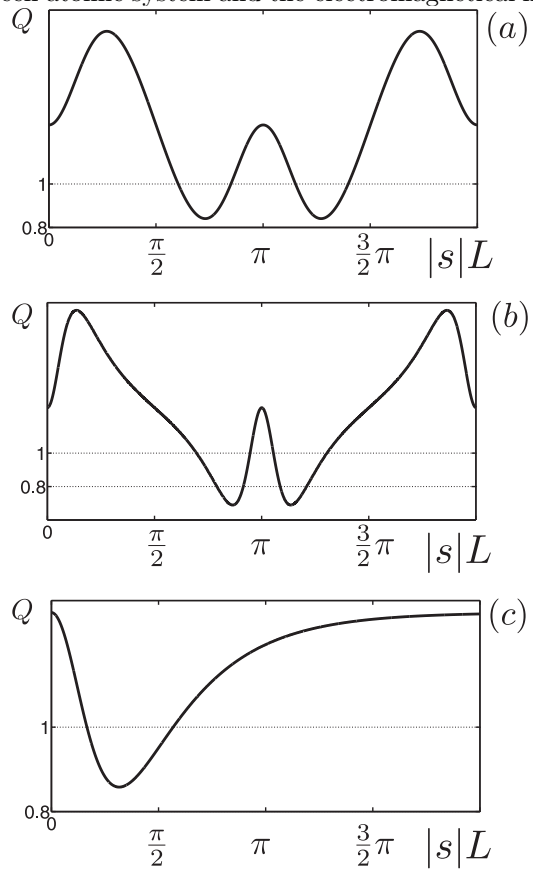}%
\caption{The function $Q$\ that is an entanglement creteria as a function of
dimentionless parameter $\left\vert s\right\vert L$ where $s$\ is given in
Eq.\ref{s} and $L$\ is the interaction length.\ a) $P=10$ b) $P=1.1$ c)
$P=0.4$}%
\label{Fig4Ent}%
\end{center}
\end{figure}

\bigskip In Fig. \ref{Fig4Ent} we plot the function $Q$\ as compared with the
entanglement criteria \cite{EntCreteria}. We see that, for $P=10$\ and
$P=1.1$, there are two dips around $\left\vert s\right\vert L=\pi+\pi l$\ for
$l=0,1,2...$, to have the two modes entangled; see Figs. \ref{Fig4Ent} (a) and
\ref{Fig4Ent} (b). And when $P=0.4$, we have only one dip where these modes
are entangled; see Fig. \ref{Fig4Ent} (c). This parametrically induced beam
splitter is capable of entangling two light fields as the usual beamsplitter
\cite{BSentanglement}.

Before ending the discussion on these applications, we give an estimation of
the experimental requirement to observe the phenomena. An ensemble of atomic
density $N\sim10^{12}$~cm$^{-3}$ and with the size $L\sim10$~cm suffices to
realize the parameters used in the above figures. One could use sodium vapor
on the $D1$ spectral line as the medium, while choosing $\Delta\sim
3000$~MHz, $\delta\sim50$~MHz, $\Omega\sim60$~MHz for the light fields.

\bigskip

\section{Conclusion}

In summary, we have studied the engineering of two quantum modes
counterpropagating in a medium of three-level atomic system also in the
presence of a strong classical standing electromagnetic field. The interaction
between atomic system and the electromagnetical modes is via dispersive FWM,
to have all fields well detuned from one-photon and two-photon resonances, so
that the FWM process is purely dispersive and the atomic system itself will be
always in the ground state $\left\vert a\right\rangle $. Under such condition
there will be no loss in the system in the presence of decays and Langevin
noises. The derived effective Hamiltonian has the similar form to a
beamsplitter Hamiltonian \cite{IntroQOGerryKnight}. We demonstrate that there
are possibilities for the parametric amplitude amplification of a squeezed
state, as well as the squeezing transferring and the entanglement generation,
with such type of interaction. Also we have found that, in a certain range of
the parameters as in Eq. (\ref{BG}), the features similar to those in a
photonic crystal band gap appear.

This work was supported by the Russian Foundation for Basic Research under
Grant No. 15-02-03959.

\end{document}